\documentclass{llncs}

\usepackage{amssymb}
\usepackage{amsmath}
\usepackage[]{algorithm2e}
\usepackage{algorithmic}
\usepackage{tikz-cd}
\usepackage{caption}
\usepackage{subcaption}
\usepackage{csquotes}

\newcommand{\dright}{\mathit{right}}
\newcommand{\dleft}{\mathit{left}}

\usetikzlibrary{decorations.pathmorphing}

\newcommand{\limp}{\Rightarrow}

\newcommand{\vp}{\vec{p}}
\newcommand{\vq}{\vec{q}}

\newcommand{\Proc}{\mathit{Proc}}

\newcommand{\Bad}{\mathit{Bad}}
\newcommand{\Init}{\mathit{Init}}

\newcommand{\cJ}{\mathcal{J}}
\newcommand{\Mod}{\mathit{Mod}}

\newcommand{\eqdef}{\triangleq}
\newcommand{\distinct}{\mathit{distinct}}
\renewcommand{\implies}{\limp}
\newcommand{\comp}{\mathit{comp}}
\newcommand{\fnext}{\mathit{next}}
\newcommand{\id}{\mathit{id}}
\newcommand{\BTW}{\mathcal{BTW}}
\newcommand{\btw}{\mathit{btw}}
\newcommand{\LO}{\mathit{LO}}

\newcommand{\Inv}{\mathit{Inv}}

\newcommand{\nbd}{\mathit{nbd}}

\newcommand{\FOL}{\mathit{FOL}}
\newcommand{\FOLC}{\FOL^\cC}
\newcommand{\CrossInit}{\mathit{CrossInit}}
\newcommand{\CrossInv}{\mathit{CrossInv}}
\newcommand{\Pred}{\mathit{Pred}}
\newcommand{\Func}{\mathit{Func}}
\newcommand{\rbr}{\mathcal{RBR}}
\newcommand{\Top}{\mathit{Top}}
\newcommand{\VC}{\mathit{VC}}
\newcommand{\cI}{\mathcal{I}}
\newcommand{\cC}{\mathcal{C}}
\newcommand{\cS}{\mathcal{S}}
\newcommand{\InvOk}{\mathit{InvOk}}
\newcommand{\Invr}{\mathit{Inv}_\mathit{red}}
\newcommand{\Invb}{\mathit{Inv}_\mathit{black}}
\newcommand{\Red}{\mathit{Red}}
\newcommand{\Black}{\mathit{Black}}

\newcommand{\fleft}{\mathit{left}}
\newcommand{\fright}{\mathit{right}}
\newcommand{\Frame}{\mathit{Frame}}

\newcommand{\sectS}{\vspace{-0.42cm}}
\newcommand{\sectP}{\vspace{-0.32cm}}
\newcommand{\picC}{\vspace{-0.32cm}}
\newcommand{\picS}{\vspace{-0.52cm}}
\newcommand{\parS}{\vspace{-.32 cm}}
\newcommand{\thmS}{\vspace{-.22 cm}}
\newcommand{\thmSz}{\vspace{-.32cm}}

\usetikzlibrary{positioning}
\usetikzlibrary{automata,arrows}
\usetikzlibrary{shapes}
\tikzset{main node/.style={circle,draw,minimum size=.5cm,inner sep=0pt}}

\let\gets\relax
\DeclareMathOperator{\gets}{:=}
\title{Local Reasoning for Paramaterized First Order Protocols}
\author{Rylo Ashmore  \and Arie Gurfinkel \and Richard Trefler }
\institute{University of Waterloo}
\begin{document}
\maketitle
\vspace{-.3cm}
\sectS
\begin{abstract}

  First Order Logic (FOL) is a powerful reasoning tool for program
  verification. Recent work on Ivy shows that FOL is well suited for
  verification of parameterized distributed systems. However,
  specifying many natural objects, such as a ring topology, in FOL is
  unexpectedly inconvenient. We present a
  framework based on FOL for specifying distributed multi-process
  protocols in a process-local manner together with an implicit network
  topology. In the specification framework, we provide an auto-active analysis technique to reason about the protocols locally, in a process-modular
  way. Our goal is to mirror the way designers often describe and reason about
  protocols. By hiding the topology behind the FOL structure, we
  simplify the modelling, but complicate the reasoning. To deal with
  that, we use an oracle for the topology 
  to develop a sound and relatively complete
  proof rule that reduces reasoning about the implicit topology back to
  pure FOL. This completely avoids the need to axiomatize the
  topology. Using the rule, we establish a
  property that reduces verification to a fixed number of processes
  bounded by the size of local neighbourhoods. We show how to use the framework on two examples, including leader election on a ring.
  


\end{abstract}

\vspace{-0.6cm}
\sectS
\section{Introduction}
\label{sec:intro}
\sectP

Auto-active\cite{Leino:2010:DAP:1939141.1939161} and automated verification engines are now commonly used to analyze the behavior
of safety- and system-critical multi-process distributed systems.  Applying the analysis techniques
early in the design cycle has the added advantage that any errors or bugs found are less
costly to fix than if one waits until the system is deployed.  Therefore, it is typical to seek a
proof of safety for \emph{parametric} designs,
where the number of participating program components is not yet determined, but the inter-process
communciation fits a given pattern, as is common in routing or communication protocols, and \mbox{other distributed systems.} 

Recently, Ivy~\cite{DBLP:conf/pldi/PadonMPSS16} has been introduced as
a novel auto-active verification technique (in the style of
Dafny~\cite{Leino:2010:DAP:1939141.1939161} for reasoning about
parameterized systems.  Ivy models protocols in First Order Logic
(FOL). The verification conditions are compiled (with user help) to a
decidable fragment of FOL, called
Effectively Propositional Reasoning (EPR)~\cite{DBLP:journals/jar/PiskacMB10}. Ivy is automatic in the sense
that the verification engineer only provides an inductive
invariant. Furthermore, unlike Dafny, it guarantees that the
verification is never stuck inside the decision procedure (verification conditions are
decidable).

One of the disadvantages of Ivy is that an engineer must formally specify the
entire protocol, including the topology. For instance, in verifying
the leader election on a ring, Ivy requires an explicit axiomatization
of the ring topology, as shown in Fig.~\ref{ringTop}. The predicate
$\btw(x,y,z)$ means that a process $y$ is between processes $x$ and
$z$ in the ring; similarly, $\fnext(a,b)$ means that $b$ is an
immediate neighbour of $a$ on the ring.  All (finite) rings satisfy
the axioms in Fig.~\ref{ringTop}. The converse is not true in
general. For instance, take the rationals $\mathbb{Q}$ and let
$\btw(x,y,z)$ be defined as $x<y<z\vee y<z<x\vee z<x<y$. All axioms of
$btw$ are satisfied, but the only consistent interpretation of
$\fnext$ is an empty set. This satisfies all the axioms, but does not
define a ring. For the axioms in Fig.~\ref{ringTop}, all
\textit{finite} models of $\btw$ and $\fnext$ describe rings. This is
not an issue for Ivy, since infinite models do not need to be
considered for EPR. Such reasoning is non-trivial and is a burden
on the verification engineer. As another example, we were not able to
come up with an axiomatization of rings of alternating red and black
nodes (shown in Fig.~\ref{fig:rbr}) within EPR. In general, a complete
axiomatization of the topology might be hard to construct.

		\begin{figure}[t]
			\begin{align*}
				\forall x,y,z&\cdot \btw(x,y,z)\implies \btw(y,z,x)\\
				\forall w,x,y,z&\cdot \btw(w,x,y)\wedge \btw(w,y,z)\implies \btw(w,x,z)\\
				\forall w,x,y&\cdot \btw(w,x,y)\implies \neg \btw(w,y,x)\\
				\forall w,x,y&\cdot \distinct(w,x,y)\implies (\btw(w,x,y)\vee \btw(w,y,x))\\
				\forall a,b&\cdot (\fnext(a,b)\iff \forall x\cdot x\neq a\wedge x\neq b\implies \btw(a,b,x))
			\end{align*}
            \picC
			\caption{A description of a unidirectional ring in FOL as presented by Ivy~\cite{DBLP:conf/pldi/PadonMPSS16}.}
			\label{ringTop}
            \picS
\end{figure}


In this paper, we propose to address this problem by specifying the
topology independently of process behaviour. We present a framework
which separates the two and provides a clean way to express the
topology. We then specify our transitions locally, as this is a
natural and common way to define protocols. Once these
preliminaries are done, we provide a process-local proof rule to verify properties
of the system. To generate the proof rule, we offload topological
knowledge to an oracle that can answer questions about the topology. Finally, we prove various properties of the proof rule.

In summary, the paper makes the following contributions. First, in
Sec.~\ref{sec:fo-prot}, we show how to model protocols locally in
FOL. This is an alternative to the global modelling used in
Ivy. Second, in Sec.~\ref{sec:fol-proof}, we show a proof rule with
verification conditions (VC) in FOL, which are often in EPR. When the VC is in EPR, this gives an engineer a mechanical check of inductiveness. This allows
reasoning about topology without axiomatizing it. Third, in
Sec.~\ref{sec:props}, we show that our proof rule (a) satisfies a
small model property, and (b) is relatively complete. The first
guarantees the verification can be done on small process domains;
the second ensures that our proof rule is fairly expressive.

We illustrate our approach on two examples.  First, as a running example, motivated by
\cite{DBLP:conf/tacas/NamjoshiT18}, is a protocol on rings of alternating red and black nodes.  These rings have only rotational
symmetry, however, they have substantial local symmetry~\cite{DBLP:conf/vmcai/NamjoshiT12,DBLP:conf/tacas/NamjoshiT16,DBLP:conf/tacas/NamjoshiT18} consisting of two equivalence classes, one of red nodes, and
one of black nodes.  Second, in Sec.~\ref{sec:le}, we consider a
modified version of the leader election protocol from Ivy
\cite{DBLP:conf/pldi/PadonMPSS16}. This is of particular interest,
since the local symmetry
of~\cite{DBLP:conf/vmcai/NamjoshiT12,DBLP:conf/tacas/NamjoshiT16,DBLP:conf/tacas/NamjoshiT18}
has not been applied to leader election. We thus
extend~\cite{DBLP:conf/vmcai/NamjoshiT12,DBLP:conf/tacas/NamjoshiT16,DBLP:conf/tacas/NamjoshiT18}
by both allowing more symmetries and infinite-state systems.



\sectS
\section{Preliminaries}
\label{sec:background}
\sectP

\paragraph{FOL syntax and semantics.}
We assume some familiarity with the standard concepts of
many sorted First Order Logic (FOL). A signature $\Sigma$
consists of sorted predicates, functions, and constants. Terms are
variables, constants, or (recursively) $k$-ary functions applied to $k$
other terms of the correct sort. For every $k$-ary predicate $P$ and
$k$ terms $t_1,\ldots,t_k$ of the appropriate sort for $P$, the formula $P(t_1,\dots,t_k)$ is a
well-formed formula (wff). Wffs are then boolean combinations of formulae and universally or existentially quantified formulae. Namely, if $\psi$ and $\varphi$ are wffs, then so are
$(\psi\wedge\varphi)$, $(\psi\vee\varphi)$,$(\neg \psi)$,
$(\psi\implies \varphi)$,$(\psi\iff\varphi)$, $(\forall x \cdot \psi)$, and
$(\exists x \cdot \psi)$. A variable $x$ in a formula $\psi$ is bound if
it appears under the scope of a quantifier. A variable not bound is free. A wff
with no free variables is called a sentence. For convenience, we often
drop unnecessary parenthesis, and use $\top$ to denote true and $\bot$ to denote false.

An FOL interpretation $\cI$ over a domain $D$ assigns every $k$-ary
predicate $P$ a sort-appropriate semantic interpretation
$\mathcal{I}(P):D^k\to\{T,F\}$; to every $k$-ary function $f$ a
sort-appropriate interpretation $\cI(f):D^k\to D$, and to every
constant $c$ an element $\cI(c) \in D$. Given an interpretation $\cI$ and
a sentence $\psi$, then either $\psi$ is true in $\cI$ (denoted,
$\cI \models \psi$), or $\psi$ is false in $\cI$ (denoted
$\cI \not \models \psi$). The definition of the models relation is
defined on the structure of the formula as usual, for example,
$\cI \models (\varphi \land \psi)$ iff $\cI \models \varphi$ and
$\cI \models \psi$.

We write $\cI(\Sigma')$ to denote a restriction of an interpretation
$\cI$ to a signature $\Sigma'\subseteq \Sigma$. Given disjoint
signatures $\Sigma$,$\Sigma'$ and corresponding interpretations $\cI$,
$\cI'$ over a fixed domain $D$, we define $\cI\oplus\cI'$ to be an
interpretation of $\Sigma\cup\Sigma'$ over domain $D$ defined such
that $(\cI\oplus\cI')(t) = \cI(t)$ if $t \in \Sigma$,
and $(\cI\oplus\cI')(t) = \cI'(t)$ if $t \in \Sigma'$. Given interpretation $\cI$ and sub-domain $D'\subseteq D$ where $D'$ contains all constants, we let $\cI(D')$ be the interpretation restricted to domain $D'$.
\parS
\paragraph{FOL modulo structures}
We use an extension of FOL to describe structures, namely graphs. In this
case, the signature $\Sigma$ is extended with some pre-defined
functions and predicates, and the interpretations are restricted to
particular intended interpretations of these additions to the
signature. We identify a structure class $\cC$ with its signature
$\Sigma_\cC$ and an intended interpretation. We write $\FOLC$
for First Order Logic over the structure class $\cC$. Common examples are
FOL over strings, FOL over trees, and other finite structures.

A structure $\cS=(D,\cI)$ is an intended interpretation
$\cI$ for structural predicates/functions $\Sigma_\cC$
over an intended domain $D$. A set of structures is denoted
$\cC$. The syntax of $\FOLC$ is given by the syntax for FOL with signature $\Sigma\uplus \Sigma_\cC$ (where $\Sigma$ is an arbitrary disjoint signature). For semantics, any $FOL$ interpretation $\cI$ of signature $\Sigma$ leads to an $FOL^\cC$ interpretation $\cI\oplus \cI_S$ of the signature $\Sigma\uplus\Sigma_\cC$. We write $\models_\cC\varphi$ iff every $FOL^\cC$ interpretation $\cI$ satisfies $\cI\models \varphi$. We introduce a process sort $\Proc$ and require the intended domain $D$ to
be exactly the set of $\Proc$-sorted elements, so that we put our intended structure
on the processes.

\parS
\paragraph{First Order Transition Systems.}
We use First Order Transitions Systems from
Ivy~\cite{DBLP:conf/pldi/PadonMPSS16,DBLP:journals/pacmpl/PadonHLPSS18}. While
the original definition was restricted to the EPR fragment of FOL, we
do not require this. A transition system is a tuple
$Tr=(S,S_0,R)$, where $S$ is a set of states, $S_0\subseteq S$ is a
set of initial states, and $R\subseteq S\times S$ is a transition
relation. A trace $\pi$ is a (finite or infinite) sequence of states
$\pi=s_0\cdots s_i\cdots$ such that $s_0\in S_0$ and for every
$0\leq i<|\pi|$, $(s_i,s_{i+1})\in R$, where $|\pi|$ denotes the
length of $\pi$, or $\infty$ if $\pi$ is infinite. A transition system may be augmented with a set
$B\subseteq S$ of \textquote{bad} states. The system is safe iff all traces
contain no bad states. A set of states $I$ is inductive iff $S_0\subseteq I$ and if $s\in I$ and $(s,s')\in R$, then $s'\in I$. Showing the existence of an inductive set $I$ that is disjoint from bad set $B$ suffices to show a transition system is safe.

A First-Order Transition System Specification (FOTSS) is a tuple
$(\Sigma,\varphi_0,\tau)$ where $\Sigma$ is an FOL signature,
$\varphi_0$ is a sentence over $\Sigma$ and $\tau$ is a sentence over
$\Sigma\uplus \Sigma'$, where $\uplus$ denotes disjoint union and  $\Sigma'=\{t'\mid t\in \Sigma\}$.  The
semantics of a FOTSS are given by First Order Transition Systems
(FOTS). Let $D$ be a fixed domain. A FOTSS $(\Sigma,\varphi_0,\tau)$
defines a FOTS over $D$ as follows:
$S=\{\cI \mid \cI\text{ is an FOL interpretation over } D\}$,
$S_0 = \{ \cI \in S \mid \cI \models \varphi_0 \}$, and
$R = \{ (\cI_1, \cI_2) \in S\times S \mid \cI_1 \oplus \cI'_2 \models \tau\}$, where $\cI'$ interprets $\Sigma'$. We may augment a FOTSS with a FOL sentence $\Bad$, giving bad states in the FOTS by $\mathcal{I}\in B$ iff $\mathcal{I}\vDash \Bad$.
A FOTSS is safe if all of its corresponding FOTS $Tr$ are safe, and is
unsafe otherwise. That is, an FOTSS is unsafe if there exists at least
one FOTS corresponding to it that has at least one execution that
reaches a bad state. 
A common way to show a FOTSS is safe is to give a formula $\Inv$ such that
 $\models \varphi_0\implies \Inv$ and
 $\models \Inv\wedge \tau\implies \Inv'$. 
Then for any FOTS over domain $D$, the set $I\subseteq S$ given by $I=\{\cI\in S\mid \cI\models \Inv\}$ is an inductive set, and $\models \Inv\implies \neg \Bad$ then suffices to show that the state sets $I,B$ in the FOTS are disjoint. Finding an invariant $\Inv$ satisfying the above proves the system safe.

\begin{example} \label{ex:fotss} Consider the following
  FOTSS:
  \begin{align*}
    \Sigma &\eqdef \{Even,+,1,var\} &
    \varphi_0 &\eqdef Even(var)\\
    \tau &\eqdef (var'=(var+1)+1)\wedge Unch(Even,+,1) &
    \Bad &\eqdef \neg Even(var)
  \end{align*}
  where $Unch(Even, +, 1)$ means that $Even$, $ + $, and $1$
  have identical interpretations in the pre- and post-states of
  $\tau$.

  Our intention is to model a program that starts with an even number
  in a variable $var$ and increments $var$ by $2$ at every
  transition. It is an error if $var$ ever becomes odd. A natural invariant to conjecture is $\Inv\eqdef Even(var)$. However, since
  the signature is uninterpreted, the FOTSS does not model our
  intention.

  For example, let $D = \{0,1,2\}$, $\cI_0(Even) = \{1, 2\}$,
  $\cI_0(1) = 1$, $\cI_0(+)(a, b) = a + b \mod 3$, and
  $\cI_0(var) = 1$. Thus, $\cI_0 \models \varphi_0$. Let $\cI_1$ be
  the same as $\cI_0$, except $\cI_1(var) = 0$. Then,
  $\cI_0 \oplus \cI'_1 \models \tau$ and $\cI_1 \models \Bad$. Thus,
  this FOTSS is unsafe.
\end{example}

One way to explicate our intention in Example~\ref{ex:fotss} is to
axiomatize the uninterpreted functions and relations in FOL as part of
$\varphi_0$ and $\tau$. Another alternative is to restrict their
interpretation by restricting the interpretation of FOL to a
particular structure. This is the approach we take
in this paper. We define a First-Order (relative to $\cC$) Transition
System Specification (FOCTSS).

We need to be able to talk about the structural objects in $\Sigma_\cC$, and so we require that every FOCTSS $(\Sigma,\varphi_0,\tau)$ be an FOTSS with $\Sigma_\cC\subseteq \Sigma$. Once we have these structural objects, any structure $(D,\cI)\in\cC$ gives a FOCTS with states $\cI$ where $\cI(\Sigma_\cC)=\cI_\cC$, initial states $\cI$ where $\cI\models \varphi_0$, transitions $(\cI_1,\cI_2)$ where $\cI_1\oplus\cI_2'\models \tau$, and bad states $\cI$ for which $\cI\models \Bad$.


\sectS
\newcommand{\cR}{\mathcal{R}}
\newcommand{\var}{\textit{var}}
\section{First-Order Protocols}
\label{sec:fo-prot}
\sectP
We introduce the notion of a \emph{First-Order
  Protocol (FOP)} to simplify and restrict
specifications in a FOTS. We choose restrictions to make our protocols
asynchronous compositions of processes over static
network topologies. Each process description is relative to its process
neighbourhood. For example, a process operating on a ring has access
to its immediate left and right neighbours, and transitions are restricted to these processes. This \mbox{simplifies the modelling}. 


We begin with formalizing the concept of a network topology. As a
running example, consider a Red-Black-Ring (RBR) topology, whose instance
with 4 processes is shown in Fig.~\ref{fig:rbr}. Processes are
connected in a ring of alternating Red and Black processes. Each process is connected to two neighbours using
two links, labelled $\dleft$ and $\dright$, respectively. From the example it is clear how to extend this topology to rings of arbitrary even size.

\newcommand{\cred}{\mathit{Red}}
\newcommand{\cblack}{\mathit{Black}}
\newcommand{\ldir}{\mathit{dir}}
\newcommand{\lkind}{\mathit{kind}}
\newcommand{\ltop}{\lkind}

To formalize this, we assume that there is a unique sort $\Proc$
for processes. Define $\Sigma^\cC=\Sigma_E^\cC\uplus \Sigma_T^\cC$ to be a
\emph{topological signature}, where $\Sigma_E^\cC$ is a set of unary
$\Proc$-sorted functions and $\Sigma_T^\cC$ is a set of distinct
$k$-ary $\Proc$-sorted predicates ($k$ fixed). Functions in $\Sigma_E^\cC$ correspond
to communication edges, such as $\dleft$ and $\dright$ in our
example. Predicates in $\Sigma_T^\cC$ correspond to classes of processes,
such as $\cred$ and $\cblack$ in our example. For simplicity, we
assume that all classes have the same arity $k$. We often omit $k$
from the signature when it is contextually clear. We are now ready to define the concept of a network topology:
\begin{definition}
  \label{def:topo}
  A \emph{network topology} $\cC$ over a topological signature
  $\Sigma^\cC$ is a collection of directed graphs $G=(V,E)$ augmented
  with an edge labelling $\ldir:E\to\Sigma^\cC_E$ and $k$-node labelling
  $\lkind:V^k\to \Sigma^\cC_T$. Given a node $p$ in a graph $G = (V, E)$
  from a network topology $\cC$, the neighbourhood of $p$ is defined as
  $\nbd(p)=\{p\}\cup \{q\mid (p,q)\in E\}$, and a neighbourhood of a tuple $\vp=(p_1,\dots, p_k)$ is defined as $\nbd(\vp)=\bigcup_{i=1}^k\nbd(p_i)$. A network topology is
  \emph{deterministic} if for every distinct pair  $q, r
  \in \nbd(p)\setminus\{p\}$, $\ldir(p, q) \neq \ldir(p, r)$. That is, each
  neighbour of $p$ corresponds to a distinct name in $\Sigma_E$.
\end{definition}

Given the signatures $\Sigma_T^\cC$ and $\Sigma_E^\cC$, the intended
interpretation of a predicate $P \in \Sigma_T^\cC$ is the set of all nodes
in the network topology labelled by $P$, and the intended
interpretation of a function $f \in \Sigma_E^\cC$ is such that $f(p) = q$
if an edge $(p, q)$ is labelled by $f$ and $f(p)=p$, otherwise.

Each graph $G$ in a network topology $\cC$ provides a possible
intended interpretation for the sort of processes $\Proc$, and the
edge and node labelling provide the intended interpretation for
predicates and functions in $\Sigma^\cC$.  

\begin{figure}[t]
  \centering
  \begin{subfigure}{.48\textwidth}
    \begin{tikzpicture}[>=stealth',shorten >=1pt,auto,node distance=2 cm, scale = 1, transform shape,
    blacknode/.style={shape=circle, draw=black, accepting},
    rednode/.style={shape=circle, draw=red},
    ]

    \node[rednode] (p0) [] {$p^2_0$};
    \node[blacknode]  [right of=p0] (p1) {$p^2_1$};
    \node[rednode] (p2) [right of=p0, below of=p0]  {$p^2_2$};
    \node[blacknode]  [below of = p0] (p3) {$p^2_3$};

    \path[->]
    (p0) edge [bend left] node [] {} (p1)
    (p1) edge [bend left] node [] {} (p2)
    (p2) edge [bend left] node [] {} (p3)
    (p3) edge [bend left] node [] {} (p0)

    (p0) edge [bend left,dashed] node [] {} (p3)
    (p1) edge [bend left,dashed] node [] {} (p0)
    (p2) edge [bend left,dashed] node [] {} (p1)
    (p3) edge [bend left,dashed] node [] {} (p2)
    ;

    \matrix [draw,below right] at (current bounding box.north east) {
      \node [blacknode,label=right:Black] {}; \\
      \node [rednode,label=right:Red] {}; \\
    };
  \end{tikzpicture}

  \caption{Red-Black-Ring of 4 process. Dashed arrows are $\dright$, and
    solid are $\dleft$.}
  \label{fig:rbr}
  \end{subfigure}\hfill%
  \begin{subfigure}{.48\textwidth}
  \begin{align*}
    Init:&\quad var\gets null\\
    Tr:&\quad  black\implies \dright.var\gets r\\
         &\quad red\implies \dright.var\gets b\\
    Bad:&\quad  red\wedge var=b
  \end{align*}
  \vspace{0.33 cm}
  \caption{A simple protocol over Red-Black-Ring topology.}
  \label{fig:proto-ex}
\end{subfigure}
\picC
\caption{An example of a topology and a protocol.}
  \label{fig:exCode}
  \picS
  \vspace{-.1cm}
\end{figure}

\begin{example}
  For our running example, consider an informal description of the
  protocol shown in Fig.~\ref{fig:proto-ex} described by a set of guarded
  commands. The protocol is intended to be executed on the RBR
  topology shown in Fig.~\ref{fig:rbr}.  Initially, all processes
  start with their state variable $var$ set to a special constant
  $null$. Then, at each step, a non-deterministically chosen process,
  sends a color to its right. Every black process sends a red color $r$, and every red
  process sends a black color $b$. It is bad if a Red process ever gets a
  black color.

  To formalize the topology, for each $n> 1$, let $G_n = (V_n, E_n)$,
  where $V_n=\{p^n_i\mid 0\leq i<2n\}$, and
  $E_n=\{(p^n_i,p^n_j)\mid |i-j|\bmod 2n = 1\}$. The edge labelling is given by $\ldir(p^n_i,p^n_j)=\dright$ if
  $j = (i+1) \mod n$ and $\dleft$ if $j= (i-1) \mod n$. Processes have  colour $\lkind(p^n_i) = \cred$ if $i$ is even, and $\cblack$ if
  $i$ is odd.  Finally, we define $\rbr=\{G_n\mid n\geq 2\}$ as the
  class of Red-Black Rings (RBR).  \qed
\end{example}
\vspace{-.2cm}
Note that any set of graphs $\mathcal{G}$ with an upper bound on the out-degree of any vertex can be given a finite labelling according to the above.

\parS
\paragraph{First-Order Protocols.} Once we have specified the topology, we want to establish how processes transition. We define the syntax and semantics of a protocol.

A protocol signature $\Sigma$ is a disjoint union of a topological
signature $\Sigma_\cC$, a state signature $\Sigma_S$, and a background
signature $\Sigma_B$. Recall that all functions and relations in
$\Sigma_\cC$ are of sort $\Proc$. All elements of $\Sigma_S$ have arity
of at least $1$ with the first and only the first argument of sort $\Proc$. Elements of
$\Sigma_B$ do not allow arguments of sort $\Proc$ at all. Intuitively,
elements of $\Sigma_\cC$ describe how processes are connected, elements
of $\Sigma_S$ describe what is true in the current state of some
process, and elements of $\Sigma_B$ provide background theories, such
as laws of arithmetic and uninterpreted functions.

For an interpretation $\cI$, and a set of processes $P\subseteq \cI(Proc)$, we
write $\cI(\Sigma_S)(P)$ for the interpretation $\cI(\Sigma_S)$ restricted to processes in $P$. Intuitively, we look only at the states of $P$ and ignore the states of all other processes. 
\vspace{-.2cm}
\newcommand{\TrLoc}{\mathit{TrLoc}}
\begin{definition}
  A \emph{First-Order Protocol} (FO-protocol) is a tuple
  $P=(\Sigma,Init(\vp),$ $Mod(p),\TrLoc(p),\cC)$, where $\Sigma$ is a
  protocol signature; $p$ is a free variable of sort $\Proc$,
  $Init(\vp)$ is a formula with $k$ free variables $\vp$ of sort $\Proc$;
  $Mod(p)$ is a set of terms
  $\{ t(p) \mid t \in \ldir(E) \} \cup \{ p \}$; $TrLoc(p)$ is a formula
  over the signature $\Sigma\cup \Sigma'$, and $\cC$ is a network
  topology. Furthermore, $Init(\vp)$ is of the form
  $\bigwedge_{P \in \Sigma_T}\left(P(\vp)\implies Init_P(\vp)\right)$, where the
  arity of $P$ is $|\vp|$, and each $Init_P$ is a formula over
  $\Sigma\setminus \Sigma_\cC$ (an initial state described without reference to topology for
  each relevant topological class); and terms of sort $\Proc$ occurring in $\TrLoc(p)$ are a subset of $\Mod(p)$.
\end{definition}

A formal description of our running example is given in Figure \ref{fig:exProto} as a FO-protocol. We define the signature including $\Sigma_\cC=\{\fleft,\fright,\Red,\Black\}$, the initial states $\Init(p)$ in the restricted form, and modification set $\Mod(p)$, where we allow processes to only write to their local neighbourhood. Next we specify two kinds of transitions, a red $t_r$ and a black $t_b$ transition. Each writes to their right neighbour the colour they expect that process to be. Each process $p$ does not change the $\var$ states of $p,\fleft(p)\in \Mod(p)$. Finally, we specify our local transitions $\TrLoc(p)$ by allowing each of the sub-transitions. Note that all process-sorted terms in $\TrLoc(p)$ are in $\Mod(p)=\{\fleft(p),p,\fright(p)\}$, and we are allowed to call on topological predicates in $\TrLoc$, finishing our specification.



Furthermore, note that the
\emph{semantic} local neighbourhood $\nbd(p)$ and the set of
\emph{syntactic} terms in $Mod(p)$ have been connected. Namely, for every edge $(p,q)\in E$, there is a term $t(p)\in Mod(p)$ to refer to $q$, and for every term $t(p)\in Mod(p)$, we will refer to some process in the neighbourhood of $p$.


\begin{figure}[t]
  \centering
  \begin{align*}
    Const&=\{null_{/0},r_{/0},b_{/0}\} \qquad
    \Func =\{left_{/1},right_{/1},var_{/1}\}\\
    \Pred &=\{Red_{/1}, Black_{/1}, =_{/2}\} \qquad
    \Sigma=(Const,\Func,\Pred)\\
    Init(p)&=(Red(p)\implies var(p)=null)\wedge (Black(p)\implies var(p)=null)\\
    Mod(p)&=\{p,right(p), left(p)\}\\
    t_r(p)&= var'(right(p))=b\wedge var'(p)=var(p)\wedge var'(left(p))=var(left(p))\\
    t_b(p)&= var'(right(p))=r\wedge var'(p)=var(p)\wedge var'(left(p))=var(left(p))\\
    TrLoc(p)&= (Red(p)\implies t_r(p))\wedge (Black(p)\implies t_b(p))
  \end{align*}
  \picC
  \caption{A FO-protocol description of the system from Fig.~\ref{fig:exCode}.}
  \label{fig:exProto}
  \picS
\end{figure}

\begin{figure}[t]
  \begin{gather*}
  \begin{aligned}
    \varphi_0&\eqdef \forall \vp\cdot Init(\vp) &
    \tau&\eqdef  \exists p\cdot TrLoc(p)\wedge Frame(p)\\
    \end{aligned}\\
  \begin{aligned}
    Frame(p)&\eqdef UnMod\land \left(\forall y\cdot y\not\in Mod(p)\implies Unch(y))\right)\\
    Unch(y)&\eqdef \left(\bigwedge_{P\in Pred_{S}}\forall \vec{v}\cdot
             P(y,\vec{v})\iff P'(y,\vec{v})\right) \wedge
           \left(\bigwedge_{f\in Func_{S}}\forall \vec{v}\cdot f(y,\vec{v})=f'(y,\vec{v})\right)\\
    UnMod&\eqdef \left(\bigwedge_{P\in Pred_{B}}\forall \vec{v}\cdot
           P(\vec{v})\iff P'(\vec{v}) \right)\land \left(\bigwedge_{f\in Func_{B}}\forall \vec{v}\cdot f(\vec{v})=f'(\vec{v})\right)
  \end{aligned}
  \end{gather*}
  \picC
  \caption{An FOTS of the protocol in Fig.~\ref{fig:exProto}.}
  \label{fig:foctssSemantics}
  \picS
\end{figure}

The semantics of a protocol $P$ are given be an FOCTSS as shown in
Fig.~\ref{fig:foctssSemantics}.
The protocol signature $\Sigma$ is the same in the FOCTSS as in the FOP. Initially, $\varphi_0$
requires that all $k$-tuples of a given topology satisfy a
topology-specific initial state. Finally, to take a transition $\tau$, some
process takes a local transition $\TrLoc(p)$ modifying states of processes that
can be described using the terms in $\Mod(p)$. $Unch(y)$ guarantees that the transition does not
affect local state of processes that are outside of
$\Mod(p)$. Finally, $UnMod$ makes all functions and predicates in the background
signature retain their interpretation during the transition. Overall, this describes a general multiprocess asynchronous protocol.


This definition of a FO-protocol places some added structure on the
notion of FOTSS. It restricts how transition systems can be
specified, which might seem like a drawback. On the contrary, the
added structure provides two benefits. First, it removes the need for
axiomatizing the network topology, since the topology is given
semantically by $\cC$. Second, the system guarantees to model
asynchronous composition of processes whose local transition relation
is given by $\TrLoc$ -- a common framework for specifying and
reasoning about protocols. 



To show safety of such a system, we will be concerned with invariants which only discuss a few processes, say $\Inv(\vp)$ where $\vp=p_1,\dots, p_k$. Then our FO-invariants will be of the form $\forall\vp\cdot \Inv(\vp)$, and substituting $\varphi_0$ into our background, we find a natural check for when a given formula is inductive:
\[\InvOk\eqdef ((\forall \vp\cdot \Init(\vp))\implies (\forall \vp\cdot \Inv(\vp)))\wedge ((\forall \vp\cdot \Inv(\vp))\wedge \tau\implies (\forall \vp\cdot \Inv'(\vp)))\]
Indeed, by unpacking definitions, one sees that $\models_\cC \InvOk$ means that every state on any trace of a FOCTS satisfies $\forall\vp\cdot \Inv(\vp)$, and thus it suffices to check that $\models_\cC \forall\vp\cdot \Inv(\vp)\implies \neg \Bad$ to prove safety. We, however, will focus on the task of verifying a candidate formula as inductive or not.

To decide if a candidate is inductive or not requires reasoning in $\FOLC$. However, reasoning about FOL extended with an
arbitrary topology is difficult (or undecidable in general). We would like to
reduce the verification problem to pure FOL. One solution is to
axiomatize the topology in FOL -- this is the approach taken by
Ivy\cite{DBLP:conf/pldi/PadonMPSS16}. Another approach is to use properties of the topology to reduce
reasoning about FO-protocols to FOL. This is similar to the use of
topology to reduce reasoning about parameterized finite-state systems
to reasoning about finite combinations of finite-state systems
in~\cite{DBLP:conf/tacas/NamjoshiT16}. In the next section, we show how this
approach can be extended to FO-protocols.

\sectS

\section{Verifying FO-Protocols using First Order Logic}
\label{sec:fol-proof}
\sectP
\sloppy In this section, we present a technique for reducing verification of
FO-protocols over a given topology $\cC$ to a decision problem in pure
FOL. We assume that we are given a (modular) inductive invariant
$\forall \vq \cdot \Inv(\vq)$ of the form
$\left(\forall \vq \cdot \bigwedge_{\Top \in
    \Sigma^\cC_T}\Top(\vq)\implies \Inv_{\Top}(\vq)\right)$. That is,
$\Inv$ has a local inductive invariant $\Inv_{\Top(\vq)}$ for each
topological class $\Top$.

Given a First-Order Protocol and candidate invariant, we want to know if  $\models_{\FOL_\cC} \InvOk$. But deciding this is hard, and so we show that deciding
validity of $\InvOk$ can be done in pure FOL using modular
verification conditions in the style of
Owicki-Gries~\cite{DBLP:journals/cacm/OwickiG76} and Paramaterized
Compositional Model Checking~\cite{DBLP:conf/tacas/NamjoshiT16}.

\newcommand{\PreInv}{\mathit{PreInv}}
\newcommand{\EqClass}{\mathit{EqClass}}
\newcommand{\Terms}{\mathit{Terms}}

The input to our procedure is a formula $\Inv_{\Top}$ over signature $\Sigma_B\uplus \Sigma_S$ for
each topological class $\Top \in \Sigma^{\cC}_T$. The VC is a
collection of sentences ensuring that for each tuple of processes
$\vq$ in a topological class $\Top$, $\Inv_{\Top}(\vq)$ is true initially, is stable
under a transition of one process in $\vq$, and is stable under
interference by any other process $p$ whose execution might affect some $q_i\in \vq$. If the VC is FOL-valid, an inductive invariant has been found. If not, there will be a local violation to inductiveness, which may correspond to a global violation.

Formally, the VC is a collection of statements of the following two forms:
\begin{gather}
\forall \vq\cdot (\CrossInit_{\Top}(\vq)\implies \Inv_{\Top}(\vq))\\
\forall p,\vq\cdot ((\CrossInv_{\Top}(\Mod(p),\vq)\wedge \tau)\implies \Inv_{\Top}'(\vq))
\end{gather}
Statements of the first form require that every local neighbourhood of $\vq$ that satisfies all appropriate initial states also satisfies $\vq$'s invariant. Statements of the second form capture both transitions where $p=q_i$ for some $i$, or process $p$ acts and modifies $q_i\in \nbd(p)$, since $p$ is quantified universally. All that remains is to formally construct the statements $\CrossInit,\CrossInv$. In order to do so, we will construct a characteristic local neighbourhood of a process tuple $\vq$. We will want to use a similar construction for both, and so generalize to the characteristic that $\vq$ must satisfy in addition to other processes given by an arbitrary set $A$.

We will say that formula $\psi$ is a valid candidate for $\chi_\Top(A,\vq)$ when it is (1) over signature $\Sigma_T\cup \Sigma_E\cup \{=\}$, (2) contains only terms $A\cup \{q_i\mid q_i\in \vq\}$, and (3) is in CNF and all literals from $\Sigma_T$ appear in positive form. Intuitively, we will want to capture when elements of $A,\vq$ satisfy various topological notions given by signature $\Sigma_E\cup \{=\}$. We also never want to force some processes to be outside of some topological class. We let $\chi_{\Top}(A,\vq)$ be the strongest candidate that satisfies $\models_\cC \forall \vq\cdot \Top(\vq)\implies \chi_{\Top}(A,\vq)$. Intuitively, $\chi$ is a formula that captures all topological knowledge derivable from the topology given that we know that $\Top(\vq)$ holds. For instance, in $\rbr$, we have $\chi_{\Red}(\emptyset, q)=\Red(q)$, while expanding this for $A=\{\fleft(p),p,\fright(p)\}$ results in the following formula.
\begin{align*}
\chi_{\Red}(&\{\fleft(p),p,\fright(p)\},q)=\Red(q)\wedge \distinct(\fleft(p),p,\fright(p))\wedge\\
& ((\Red(\fleft(p))\wedge \Black(p)\wedge \Red(\fright(p))\wedge p\neq q)\vee\\& (\Black(\fleft(p))\wedge \Red(p)\wedge \Black(\fright(p))\wedge \distinct(\fleft(p),\fright(p),q)))
\end{align*}

\begin{figure}[t]
    \centering

\begin{tikzpicture}[>=stealth',shorten >=1pt,auto,node distance=1.2 cm, scale = 1, transform shape,
blacknode/.style={shape=circle, draw=black, accepting},
rednode/.style={shape=circle, draw=red},
]

\node[rednode] (p0) [] {$p$};
\node[rednode] (q0) [left =2.5 cm of p0] {$q$};
\node[blacknode]  [right of=p0] (p1) {$l(p)$};
\node[blacknode] (p3) [below of=p0]  {$r(p)$};
\node[rednode] (p2) [below of =p1] {$q$};
\node[blacknode] (p4) [right =1.5cm of p1] {$p$};
\node[rednode]  [right of=p4] (p5) {$l(p)$};
\node[rednode] (p7) [below of=p4]  {$r(p)$};
\node[rednode] (p6) [below of =p5] {$q$};

\path[->]
(p0) edge [bend left] node [] {} (p1)

(p0) edge [bend left,dashed] node [] {} (p3)
(p4) edge [bend left] node [] {} (p5)

(p4) edge [bend left,dashed] node [] {} (p7);
\path[-]
(p2) edge[dashed] (p0)
(p6) edge[dashed] (p5)
(p7) edge[dashed] (p6)
;

\matrix [draw,below right = 1cm] at (current bounding box.north east) {
    \node [blacknode,label=right:Black] {}; \\
    \node [rednode,label=right:Red] {}; \\
};
\end{tikzpicture}
\picC
    \caption{Characteristics $\chi_{\Red}(\emptyset,q)$ and $\chi_{\Red}(\Mod(p),q)$ for the $\rbr$ topology. }
    \label{fig:chiEx}
    \picS
\end{figure}
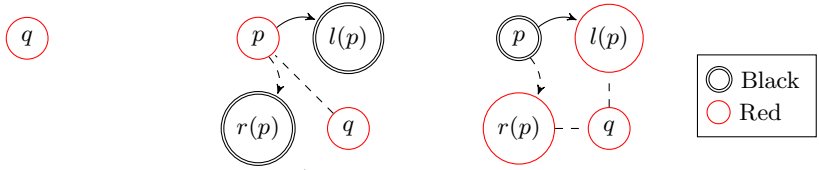

These characteristics are illustrated in Figure \ref{fig:chiEx}. When we just look at $\chi_\Red(\emptyset,q)$, we find $q$ is red. However, if we expand our local reasoning to the characteristic $\chi_\Red(\Mod(p),q)$, we find that there are two options given by $\rbr$. One option is $p$ is red, and $q=p$ is optional (dotted lines), while $q\neq \fleft(p),\fright(p)$. Alternatively, $p$ is black, and $q\neq p$, but $q$ could be $\fleft(p),\fright(p)$, or neither.

Once we have $\chi_{\Top}(A,\vq)$, we can define our statements $\CrossInit_\Top,$ $\CrossInv_\Top$. First, $\CrossInit_\Top(\vq)$ is obtained from $\chi_{\Top}(\emptyset, \vq)$ by replacing every instance of $\Top_i'(\vq)$ with $\Init_{\Top_i}'(\vq)$. We build our interference constraints in a similar way. We construct $\CrossInv_\Top(\vq)$ by modifying $\chi_\Top(\Mod(p),\vq)$. Namely, we obtain $\CrossInv_{\Top}(\Mod(p),\vq)$ from $\chi_{\Top}(\Mod(p),\vq)$ by replacing every instance of $\Top_i'(\vq)$ with $\Top_i'(\vq)\wedge \Inv_{\Top_i}'(\vq)$.

\begin{example}
  The VC generated by the $\rbr$ topology may be partitioned into $\VC_\Red$ and $\VC_\Black$, each consisting of the statements whose conclusions are $\Inv_\Red,\Inv_\Red'$ and $\Invb,\Invb'$, respectively. $\VC_\Red$ is shown
  in Fig.~\ref{fig:rbrVC}. The conditions for $\VC_\Black$ are symmetric. One can check that
  \begin{align*}
    \Invr(p) &\eqdef \var(p) \neq b &
    \Invb(p) &\eqdef \top
  \end{align*}
  is an inductive invariant for the protocol in Fig.~\ref{fig:exCode}.
  \qed
\end{example}

\begin{figure}[t]
  \centering
  \begin{gather}
    \forall p\cdot \Init_{red}(p)\implies \Inv_{red}(p)\label{eq:initred}\\ \hline \nonumber
    \forall p,q\cdot (\Red(q)\wedge \Invr(q)\wedge \Red(\fleft(p))\wedge \Invr(\fleft(p))\wedge\nonumber\\
     \Black(p)\wedge \Invb(p)\wedge\Red(\fright(p))\wedge \Invr(\fright(p))\wedge\nonumber\\ p\neq q\wedge \distinct(\fleft(p),p,\fright(p)))\implies \Invr'(q)
\label{eq:black}\\
    \forall p,q\cdot (\Red(q)\wedge \Invr(q)\wedge \Black(\fleft(p))\wedge\Invb(\fleft(p))\wedge\nonumber\\ \Red(p)\wedge \Invr(p)\wedge \Black(\fright(p))\wedge \Invb(\fright(p))\wedge\nonumber
    \\ \distinct(\fleft(p),\fright(p),q)\wedge \distinct(\fleft(p),p,\fright(p)))\implies \Invr'(q)
\label{eq:vcExEnd}
    \\ \hline
    VC_{P,1}(\Invr,\Invb)\eqdef \eqref{eq:initred} \land
      \eqref{eq:black}\land \eqref{eq:vcExEnd}
  \end{gather}
  \picC
  \caption{The verification conditions $\VC_\Red$ for the red process invariant.}
  \label{fig:rbrVC}
  \picS
\end{figure}

In practice, the role of the oracle can be filled by a verification engineer. A description of local neighbourhoods starts by allowing all possible neighbourhoods, and as counter-examples disallowed by the topology occur, a verifier may dismiss local configurations that cannot occur on the topology.


\sectS
\section{Soundness and Completeness}
\label{sec:props}
\sectP
In this section, we present soundness and relative completeness of our
verification procedure from Section~\ref{sec:fol-proof}.
\parS
\paragraph{Soundness.} To show soundness, we present a model-theoretic
argument to show that whenever the verification condition from
Section~\ref{sec:fol-proof} is valid in FOL, then the condition
$\InvOk$ is valid in FOL extended with the given topology $\cC$.
\thmS
\vspace{.1cm}
\begin{theorem}
  \label{thm:sound}
  \mbox{Given a FO-protocol $P$ and a local invariant per topological class}
  $\Inv_{\Top_1}(\vp),\ldots,\Inv_{\Top_n}(\vp)$, if $\vDash VC(\Inv)$,
  then $\vDash_\mathcal{C} InvOk(\Inv)$.
\end{theorem}
\thmSz
\begin{proof}
  We show that $\InvOk(\Inv)$ is valid in $\FOLC$ by showing that
  any pair of $\FOLC$ interpretations $\cI$ and $\cI'$ satisfy
  $VC(\Inv)$ as FOL interpretations, and this is strong enough to guarantee
  $\cI\oplus \cI'\models \InvOk(\Inv)$.

  Let $\cI,\cI'$ be $FOL^\cC$ interpretations over some
  $G=(V,E)\in \cC$. Then $\cI\oplus\cI'\models VC(\Inv)$ because
  $VC(\Inv)$ is valid and $\cI \oplus\cI'$ is an FOL interpretation.

  We first show that
  $\cI\models (\forall \vp\cdot \Init(\vp)\implies \forall \vp \cdot
  \Inv(\vp))$. Suppose that $\cI\vDash \forall \vp\cdot
  \Init(\vp)$. Let $\vp$ be an aribtrary tuple in $G$. If $\cI\models \neg \Top_i(\vp)$ for every $\Top_i\in \Sigma_T$, then $\Inv(\vp)$ follows vacuously. Otherwise, suppose $\cI\models \Top_i(\vp)$. Then by definition of $\chi$, we obtain $\cI\models \chi_{\Top_i}(\emptyset,\vp)$ since $\cI\models \Top_i(\vp)\implies \chi_{\Top_i}(\vp)$. Since $\cI\models \forall \vp\cdot \Init(\vp)$, this gives us that $\cI\models \CrossInit(\vp)$ (for any $\Top_j(\vp')$ in $\chi_{\Top_i}(\vp)$, find that $\Init(\vp')$, and thus $\Top_j(\vp')$ implies $\Init_{\Top_j}(\vp')$, giving $\CrossInit$). Since $\cI\models \CrossInit_{\Top_i}(\vp)$ and $\cI\models VC$, we get $\cI\models \CrossInit_{\Top_i}(\vp)\implies \Inv_{\Top_i}(\vp)$, finally giving us $\cI\models \Inv_{\Top_i}(\vp)$, as desired.


  %

  Second, we show that
  $\cI\oplus \cI'\models (\forall \vp\cdot \Inv(\vp))\wedge
  \tau\implies (\forall \vp\cdot \Inv(\vp))$. Suppose that
  $\cI\models \forall \vp\cdot \Inv(\vp)$ and
  $\cI\oplus \cI'\models \TrLoc(p)\wedge \Frame(p)$ for some
  $p\in V$. We show that $\cI'\models \forall \vq\cdot
  \Inv'(\vq)$. Let $\vq\in V^k$ be an arbitrary process tuple. If
  $\cI'\not\models \Top_i(\vq)$ for all $1\leq i\leq n$, then
  $\cI'\models \Inv'(\vq)$ vacuously. Suppose
  $\cI'\models \Top_i(\vq)$ for some $\Top_i\in \Sigma_T$. Then $\cI\models \Top_i(\vq)\implies \chi_{\Top_i}(\Mod(p),\vq)$, and so $\cI\models \chi_{\Top_i}(\Mod(p),\vq)$. Again by instantiating $\forall \vp\cdot \Inv(\vp)$ on terms in $\Mod(p),\vq$, we may obtain that $\cI\models \CrossInv(\Mod(p),\vq)$. Combined, we have $\cI\oplus \cI'\models \CrossInv(\Mod(p),\vq)\wedge \tau$. Applying $VC$ finally gives $\Inv_{\Top_i}(\vq)$.\qed


\end{proof}

Intuitively, the correctness of Theorem~\ref{thm:sound} follows from
the fact that any interpretation under $\FOLC$ is also an
interpretation under $\FOL$, and all preconditions generated for VC
are true under $\FOLC$ interpretation.
\parS
\paragraph{Small model property.} Checking validity of universally quantified statements in FOL is in the fragment EPR, and thus we obtain a result saying that we only need to consider models of a given size. This means that a FOL solver needs to only reason about finitely many elements of sort $\Proc$. It further means that topologies such as $\rbr$ may be difficult to compile to EPR in Ivy, but our methodology guarantees our verifications will be in EPR.
\thmS
\vspace{.1cm}
\begin{theorem}
    If $\models VC(\Inv)$ for all process domains of size at most
    $|\Mod(p)|+k$, then $\models_\cC \InvOk(\Inv)$.
\end{theorem}
\thmSz
\begin{proof}
    By contrapositive, suppose $\not\vDash_\mathcal{C}
    InvOk(\Inv)$. Then, by Theorem~\ref{thm:sound},
    $\not\models VC(\Inv)$. Let $\cI \oplus \cI'$ be a falsifying
    interpretation. It contains an assignment to $\Mod(p)$ and $\vq$, or
    to $\vp$ that makes at least one statement in $VC(\Inv)$
    false. Then, $(\cI \oplus \cI')(\Mod(p) \cup \vq)$ or $\cI(\vp)$
    $\Mod(p),\vec{q}$ or $\vp$ to make some statement in $VC$
    unsatisfiable. Then
    $\mathcal{I}\oplus \mathcal{I}'(\Mod(p)\cup \vq)$ or $\cI(\vp)$ is
    also a counter-model to $VC(\Inv)$, but with at most
    $|Mod(p)|+k$ elements of sort $\Proc$.
\end{proof}
\parS
\paragraph{Relative Completeness.}
We show that our method is relatively complete for local invariants
that satisfy the \emph{completability} condition. Let $\varphi(\vp)$ be a
formula of the form
$\bigwedge^n_{i=1}(\Top_i(\vp)\implies \varphi_{\Top_i}(\vp))$ with
$\varphi_{\Top_i}(\vp)$ over the signature $\Sigma_S\cup
\Sigma_B$. Intuitively, $\varphi(\vp)$ is \emph{completable} if every
interpretation $\cI$ that satisfies $\forall \vp \cdot \varphi(\vp)$
and is consistent with some $\cC$-interpretation $\cI_G$ can be
extended to a full $\cC$-interpretation (not necessarily $\cI_G$) that
satisfies $\forall\vp\cdot\varphi(\vp)$.  Formally, $\varphi$ is
\emph{completable} relative to topology $\cC$ iff for every
interpretation $\cI$ with domain $U\subseteq V$ for $G=(V,E)\in \cC$
with an intended interpretation $\cI_G$ such that
$(\cI\uplus \cI_G)(U)\models \forall \vp\cdot \varphi(\vp)$, there
exists an interpretation $\cJ$ with domain $V$ s.t.
$(\cJ\uplus \cI_G)\models \forall \vp\cdot \varphi$ and
$\cI(U)=\cJ(U)$.  Furthermore, we need a lemma expressing that the
chacteristic formula $\chi_\Top(A,\vq)$ captures all the information we care about for a
given topological neighbourhood of $A,\vq$.
\thmS
\begin{lemma}\label{lem:THElemma}

    If FOL interpretation $\cI$ of signature $\Sigma_\cC$ satisfies $\cI\models \chi_{\Top}(A,\vq)$, then there exists a $\cC$ interpretation $\cJ$ of the same signature with $\cJ\models \chi_\Top(A,\vq)$ and $\cI\models t_i=t_j$ iff  $\cJ\models t_i=t_j$ for terms $t_i,t_j\in A\cup \vq$.
\end{lemma}

\thmSz
\begin{proof}
  Let $\cI\models
  \chi_{\Top}(A,\vq)$. 
  Let $\varphi(A,\vq)$ be the conjunction of all atomic formulae over
  the signature $\{=\}$ and statements $\neg \Top_j(\vq')$ that is
  true of elements of $A,\vq$ in interpretation $\cI$. If no $\cC$
  interpretation $\cJ\models \Top(\vq)\wedge\varphi(A,\vq)$, then we
  can add the clause $\neg \varphi(A,\vq)$ to $\chi_\Top(A,\vq)$, thus
  strengthening it (this is stronger since
  $\cI\models \Top(\vq), \not\models \neg\varphi(A,\vq)$, and is true
  of every interpretation modelling $\Top(\vq)$). However, this
  violates the assumptions that $\chi_\Top$ is as strong as possible. Thus, some $\cJ\models \Top(\vq)\wedge
  \varphi(A,\vq)$. Note that $\cJ$ already satisfies $t_i=t_j$ iff $\cI$ satisfies $t_i=t_j$ since every statement of $=,\neq$ is
  included in $\varphi(A,\vq)$. Finally, since $\cJ$ is a $\cC$
  interpretation and $\cI'\models \Top(\vq)$, then
  $\cI'\models \chi_\Top(A,\vq)$ by definition.\qed
\end{proof}

\thmS
\vspace{-.2cm}
\begin{theorem}
  \label{thm:complete}
  Given an FO-protocol $P$, if $\models_\cC InvOk(\Inv)$ and both
  $Inv(\vec{p})$ and $Init(\vp)$ are completable relative to $\cC$,
  then $\models VC(\Inv)$.
\end{theorem}
\thmSz
\newcommand{\cM}{\mathcal{M}}
\begin{proof}
  By contra-positive, we show that given a completable local invariant
  $\Inv(\vp)$, if $VC(\Inv)$ is falsifiable in $\FOL$, then
  $\InvOk(\Inv)$ is falsifiable in $\FOLC$. Suppose $VC(\Inv)$ is not valid, and let $\cI \oplus \cI'$ by such
  that $\cI\oplus\cI'\not\models VC(\Inv)$. We consider two cases -- a violation initially or inductively.

  \emph{Case 1: Initialization:} For some processes
  $\vp = \langle p_1, \ldots, p_k \rangle $ and
  $1 \leq i \leq |\Sigma^\cC_T|$,
  $\cI \models \CrossInit_{\Top_i}(\vp)$ and $\cI \not\models \Inv_{\Top_i}(\vp)$.
  Modify $\cI(\Sigma_T)$ for every $\vq$ so that $\Top_j(\vq)$ is
  interpreted to be true iff $\Init_{\Top_j}(\vq)$ is true. Noting
  that all invariants and initial conditions are outside of the
  signature $\Sigma_T$, we observe that this is done without loss of
  generality. Since $\cI\models \CrossInit_{\Top_i}(\vp)$, we conclude
  now that $\cI\models \chi_{\emptyset, \Top_i}(\vp)$. Applying Lemma \ref{lem:THElemma} to $\cI(\Sigma_\cC)$, we get a $\cC$ interpretation
  $\cJ\models \chi_{\Top_i}(\emptyset, \vp^\cC)$. Since this model has
  the same equalities of terms $\vp^\cC$ in $\cJ$ as $\vp$ in $\cI$, we may copy the states $\cI(\Sigma_S)(p_i)$ to $\cJ(\Sigma_S)(p_i)$. Set $\cJ(\Sigma_B)=\cI(\Sigma_B)$. Since $\Init$ is completable by assumption, we complete $\cJ(\Sigma_S\cup \Sigma_B)(\vp)$ to $\cJ(\Sigma_S\cup \Sigma_B)$, completing our construction of $\cJ$ interpreting $\Sigma_\cC\cup \Sigma_S\cup \Sigma_B$. Note that $\cJ\models \forall \vp\cdot \Init(\vp)$, but $\cJ\models \neg \Top_i(\vp^\cC)\wedge \Inv_{\Top_i}(\vp^\cC)$, thus showing that $\InvOk(\Inv)$ is falsifable in $\FOLC$.

%
%

  \emph{Case 2: Inductiveness:} For some $p,\vq$, and
  $1\leq i\leq |\Sigma^\cC_T|$, we have
  $\cI\models \CrossInv_{\Top_i}(\Mod(p),\vq)$,
  $(\cI\oplus\cI')\models \TrLoc(p)\wedge \Frame(p)$, and
  $\cI'\not\models \Inv_{\Top_i}(\vq)$. By construction,
  $\models\CrossInv(\Mod(p),\vq)\implies
  \chi_{\Top_i}(\Mod(p),\vq)$. Applying Lemma \ref{lem:THElemma} to
  $\cI(\Sigma_\cC)\models \chi_{\Top_i}(\Mod(p),\vq)$, we get a $\cC$
  interpretation of $\Sigma^\cC_T$ which
  $\cJ\models \chi_{\Top_i}(\Mod(p^\cC),\vq^\cC)$. We extend this
  to a full model $\cJ$ of signature
  $\Sigma_\cC\cup \Sigma_S\cup\Sigma_B$. We set
  $\cJ'(\Sigma_\cC)=\cJ(\Sigma_\cC)$. Then, since $\cJ$ and $\cI$, and
  $\cJ'$ and $\cI'$ share equalties across terms in $\Mod(p)\cup \vq$ and
  $\Mod(p^\cC) \cup \vq^\cC$, we can lift states from terms
  $t\in \Mod(p)\cup \vq$ by
  $\cJ(\Sigma_S\cup \Sigma_B)(t^\cC)\eqdef \cI(\Sigma_S\cup
  \Sigma_B)(t)$ and $\cJ'(\Sigma_S)(t^\cC)\eqdef
  \cI'(\Sigma_S)(t)$. Since $\Inv$ is completable, we complete this
  interpretation with $\cJ(\Sigma_S\cup \Sigma_B)$ and clone the
  completion to $\cJ'(\Sigma_S\cup \Sigma_B)(V\setminus (\Mod(p)\cup
  \vq))$. Overall, this completes the interpretations
  $\cJ\oplus \cJ'$.

  Note that $\cJ\models \forall \vp\cdot \Inv(\vp)$ by
  construction. Similarly, $\cJ\oplus \cJ'\models \tau$ since
  $\cI\oplus \cI'\models \tau(p)$ and $\Mod(p)$ terms are lifted
  directly from $\cI$ and $\cI'$ to $\cJ$ and $\cJ'$. Finally,
  $\cJ'\models \neg \Inv_{\Top_i}(\vq)$ since $\cJ'(\Sigma_S)$ is
  lifted directly from $\cI'(\Sigma_S)$, which is the language of
  invariants. Thus, we have shown that $\InvOk(\Inv)$ is falsifiable in
  $\FOLC$ in this case as well.\qed

\end{proof}

How restrictive is the requirement of \emph{completability}?
Intuitively, if a protocol is very restrictive about how processes
interact, then the system is likely sufficiently intricate that trying
to reason locally may be difficult independant of our methodology. For instance, the invariant we later find for leader election is not completable. However, if equivalence classes are small, then most reasonable formulae satisfy the completability condition.

\thmS
\begin{theorem}\label{thm:k1thm}
  If $\Inv_{\Top_i}(p)$ is satisfiable over any domain for each
  $1 \leq i \leq n$ and topological predicates are of arity $k=1$,
  then $\Inv(p)$ is completable.
\end{theorem}\thmSz
\begin{proof}
  Let $\Inv_i(p)$ be satisfiable for each $1\leq i\leq n$. Then let
  $\cI(V')$ be an interpretation of $\Sigma_B\uplus \Sigma_S$ over
  domain $V'\subseteq V$ for $G=(V,E)\in \cC$. For each
  $p\in V\setminus V'$, suppose $\cI_G\models \Top_i(p)$ for some
  $1\leq i\leq n$. Then choose $\cJ(p)\models \Inv_{\Top_i}(p)$ since
  $\Inv_{\Top_i}(p)$ is satisfiable. Otherwise, if
  $\cI_G\not\models \Top_i(p)$ for all $1\leq i\leq n$, then $\cJ(p)$
  is chosen arbitrarily. In either case, $\cJ\models
  \Inv(p)$. Finally, define $\cJ(p)=\cI(p)$ for $p\in V'$. Then $\cJ$
  completes the partial interpretation $\cI$.
\end{proof}

Theorem~\ref{thm:k1thm} can be generalized to the case where the topological
kinds $\Sigma_T$ are non-overlapping, and individually completable, where by individually completable, we mean that if $\Top(\vp)$ and process states of $\vp'\subset\vp$ are given, then there is a way to satisfy $\Inv(\vp)$ without changing the states of $\vp'$.

\sectS
\section{Example: Leader Election Protocol}
\label{sec:le}
\sectP
In this section, we illustrate our approach by applying it to the
well-known leader election
protocol~\cite{Chang:1979:IAD:359104.359108}. This is essentially the same protocol used to illustrate Ivy
in~\cite{DBLP:conf/pldi/PadonMPSS16}. The goal of the protocol is to
choose a leader on a ring. Each process sends messages to its neighbour on one side
and receives messages from a neighbour on the other side. Initially, all
processes start with distinct identifiers, $\mathit{id}$, that are
totally ordered. Processes pass $\id$s around the ring and declare themselves the leader if they ever receive their own $\id$. 

We implement this behaviour by providing each process a comparison variable $\comp$. Processes then pass the maximum between $\mathit{id}$ and $\mathit{comp}$ to the next process. A process whose
$\mathit{id}$ and $\mathit{comp}$ have the same value is the
leader. The desired safety property is that there is never more than
one leader in the protocol.

In~\cite{DBLP:conf/pldi/PadonMPSS16}, the protocol is modelled by a
global transition system. The system maintains a bag of
messages for each process. At each step, a currently waiting message is selected and
processed according to the program of the protocol. The network topology
is axiomatized, as shown in Section~\ref{sec:intro}.  Here, we present
a local model of the protocol and its corresponding verification
condition.
\parS
\paragraph{Network topology.} The leader election protocol operates on
a ring of size at least 3. For $n \geq 3$, let $G_n=(V_n,E_n)$, where
$V_n=\{p^n_i\mid 0\leq i<n\}$ and
$E_n=\{(p^n_i,p^n_j)\mid 0\leq i<n, j = i+1\bmod n\}$. Let
$\Sigma_E=\{\fnext\}$ and $\Sigma_T=\{\btw\}$, where $\btw$ is a
ternary relation such that $\btw(p^n_i,p^n_j,p^n_k)$ iff $i<j<k$,
$j<k<i$, or $k<i<j$. Finally, the network topology is
$\BTW=\{G_n\mid n\geq 3\}$. Note that while $\BTW$ can
be axiomatized in FOL, we do not require such an axiomatization. The
definition is purely semantic, no theorem
prover sees it.

\begin{figure}[t]
  \centering
  \begin{gather*}
    \begin{aligned}
      Const&\eqdef\{0_{/0}\} &
      \Func &\eqdef\{\fnext_{/1},\id_{/1},\comp_{/1}\} &
      \Pred &\eqdef\{\leq_{/2},=_{/2},\btw_{/3}\} &
      \cC &\eqdef \BTW\\
    \end{aligned}\\
    \begin{aligned}
      \Sigma&\eqdef(Const,\Func,\Pred) &
      \LO_0(\leq)&\eqdef\LO(\leq)\land \forall x\cdot 0\leq x &
      \Mod(p)&\eqdef\{p,\fnext (p) \}
      \\
    \end{aligned}\\
    \begin{aligned}
      \Init(p)&\eqdef\left(\LO_0(\leq) \land \btw(x,y,z)\limp(\distinct(\id(x),\id(y),\id(z))\land 0 < id(x)\wedge\comp(x)=0\right))\\
      \tau_1(p)&\eqdef\left(\id(p)\leq \comp(p)\implies \left(\comp'(\fnext(p))=\comp(p)\right)\right)\\
      \tau_2(p)&\eqdef\left(\comp(p)\leq \id(p)\implies \left(\comp'(\fnext(p))=\id(p)\right)\right)\\
      \TrLoc(p)&\eqdef \left(\id(p)=\id'(p)\land \comp(p)=\comp'(p)\land
        \id'(\fnext(p))=\id(\fnext(p))\land \tau_1(p) \land \tau_2(p)\right)
    \end{aligned}
  \end{gather*}
\picC
  \caption{A model of the Leader Election protocol as a FO-protocol.}
  \label{fig:leaderEx}
  \picS
\end{figure}

A formal specification of the leader election as an FO-protocol is
shown in Fig.~\ref{fig:leaderEx}, where $LO(\leq)$ is an
axiomatization of total order from~\cite{DBLP:conf/pldi/PadonMPSS16},
and $x < y$ stands for $x \leq y \land x \neq y$. The model follows
closely the informal description of the protocol given above. The
safety property is $\neg Bad$, where
$Bad = btw(x,y,z)\land id(x)=comp(x)\land id(y)=comp(y)$. That is, a bad state is reached when two processes that participate in the
$\btw$ relation are both leaders.

A local invariant $Inv_{\textit{lead}}$ based on the invariant
from~\cite{DBLP:conf/pldi/PadonMPSS16} is shown in
Fig.~\ref{fig:leader-inv}. The invariant first says if an $\id$ passes from $y$ to $x$ through $z$, then it must witness $\id(y)\geq \id(z)$ to do so. 
Second, the invariant says that if a process is a leader, then it has a maximum id. Finally, the invariant asserts our safety property.
\begin{figure}[t]
\centering
  \begin{gather*}
    (\btw(x,y,z) \land \id(y)=\comp(x)) \implies (\id(z)\leq \id(y)) \\
    (\btw(x,y,z) \land \id(x)=\comp(x)) \implies (\id(y)\leq \id(x)\wedge \id(z)\leq \id(x))\\
    (\btw(x,y,z) \land \id(x)=\comp(x)\land \id(y)=\comp(y)) \implies x= y
  \end{gather*}
\picC
  \caption{Local inductive invariant $Inv_{\textit{lead}}(x,y,z)$ for
    Leader Election from Fig.~\ref{fig:leaderEx}.}
\label{fig:leader-inv}
\picS
\end{figure}
%
%




This invariant was found interactively with Ivy by seeking local violations to the invariant. Our protocol's $\btw$ is uninterpreted, while Ivy's $\btw$ is explicitly axiomatized. The inductive check assumes that the processes $p,\fnext(p),\vq$ all satisfy a finite instantiation of the ring axioms (this could be done by the developer as needed if an axiomatization is unknown, and this is guaranteed to terminate as there are finitely many relevant terms), and $\btw(\vq)$. Once the invariants are provided, the check of inductiveness is mechanical. Overall, this presents a natural way to model protocols for engineers that reason locally.

\parS

        \paragraph{An uncompletable invariant} The invariant for the leader election is not completable. To see this, we present a partial interpretation $\cI$ over $\{p^3_0,p^3_2\}\subseteq V_3$ from $G_3$ with no extension. We choose $\cI(\leq)$ to be $\leq$ over $\mathbb{N}$, as intended. Then we choose $\cI(\id)$ to map $p^3_0\mapsto 1$ and $p^3_2\mapsto 2$. We also choose $\cI(\comp)$ to map $p^3_0\mapsto 0$ and $p^3_2\mapsto 1$. Since no tuple satisfies $btw$, this vacuously satisfies all invariants thus far. Let $\mathcal{J}$ be an $\BTW$ interpretation agreeing on $p^3_0,p^3_2$. Consider $\id(p^3_1)$. We know $\id(p^3_1)\neq 0,1,2$ since we require distinct ids across the new $\btw$ relation. But we also have $\id(p^3_0)=\comp(p^3_2)$ and thus to satisfy $\Inv$ we must have $\id(p^3_0)\geq \id(p^3_1)$. Thus we seek an $n\in\mathbb{N}$ such that $1\geq n$, but $n\neq 0,1$, which cannot exist. Thus $\Inv$ is uncompletable.



\sectS
\section{Related Work}
\label{sec:related}
\sectP
We have shown how  analysis techniques for parametric distributed systems composed
of several components
running on locally symmetric topologies, introduced in \cite{DBLP:conf/vmcai/NamjoshiT12,DBLP:conf/vmcai/NamjoshiT13,DBLP:conf/tacas/NamjoshiT15,DBLP:conf/tacas/NamjoshiT16,DBLP:conf/tacas/NamjoshiT18}, can be generalized and applied within a First Order Logic
based theorem proving engine. 
The key steps are showing how the local neighbourhoods of network nodes
can be encoded and reasoned about 
separately from
FOL based reasoning about safety properties of the reachable states of a parametric family of programs.
Using separate decision procedures, one for answering questions about the topology and one
for answering questions about local similarity between topological structures, and processes, one can then reason
compositionally within an FOL-based theorem prover.  We have used the framework to reason
about the correctness of a version of the leader election protocol~\cite{Chang:1979:IAD:359104.359108}.

We based our description of leader election on that presented in the
Ivy framework~\cite{DBLP:conf/pldi/PadonMPSS16}.  The distinction being
that the analysis carried out in Ivy~\cite{DBLP:conf/pldi/PadonMPSS16} is global, while the
analysis given in the current paper is local, where the local structures reason about
triples of processes in the the ring.

There has been extensive work on proving properties of parametric, distributed
protocols.  In particular the work in~\cite{Abdulla2016} offers an alternative approach
to parametric program analysis based on ``views''.   In that work, cut off points are
calculated during program analysis. As another example, in~\cite{DBLP:conf/vmcai/NamjoshiT12,DBLP:conf/tacas/NamjoshiT16,DBLP:conf/tacas/NamjoshiT18}
the ``cut-offs'' are based on the program topology and the local structural symmetries amongst the nodes
of the process interconnection networks.

The notion of a ``cutoff'' proof of safety for a parametric family of
programs was first introduced by~\cite{Emerson:1995:RR:199448.199468}. For example, in \cite{Emerson:1995:RR:199448.199468}, if a ring of 3 processes
satisfies a parametric property then the property must hold
for all rings with at least three nodes.  The technique
used here is somewhat different; rather than needing to check a ring
of 3 processes, we check all pseudo-rings of a given size.

Local symmetry reduction for multi-process networks and parametric
families of networks generalizes work on ``global'' symmetry reduction
introduced by~\cite{Emerson:1996:SMC:235947.235954}
and~\cite{Clarke:1996:EST:235947.235952}. Local symmetry is, in
general, an abstraction technique that can offer exponentially more
reduction than global symmetry. In particular, ring
structures are globaly rotationally symmetric, but for
isomorphic processes may be fully-locally symmetric~\cite{DBLP:conf/tacas/NamjoshiT16,DBLP:conf/tacas/NamjoshiT18}.

Recent work~\cite{Taube:2018:MDD:3192366.3192414} has
focused on \textit{modular} reasoning in the proof or analysis of
distributed systems.  In the current work, the modularity in the proof
is driven by a natural modularity in the program structures.  In
particular, for programs of several processes proofs are structured by
modules that are local to a neighborhood of one or more processes
\cite{DBLP:conf/vmcai/NamjoshiT12,DBLP:conf/tacas/NamjoshiT16,DBLP:conf/tacas/NamjoshiT18}.


\sectS
\section{Conclusion}
\label{sec:conclusion}
\sectP

We have presented a framework for specifying protocols in a
process-local manner with topology factored out. We show that
verification is reducible to FOL with an oracle to answer local
questions about the topology. This reduction results in a decidable VC when the background theories are decidable. This cleanly separates the reasoning
about the topology from that of the states of the
processes.

Many open questions remain.  We plan to investigate our methodology on
other protocols and topologies, implement oracles for common
topologies, and explore complexity of the generated characteristic
formulae. Finally, we restricted ourselves to static topologies of
bounded degree. Handling dynamic or unbounded  topologies,
for example \mbox{in the AODV protocol~\cite{DBLP:conf/forte/NamjoshiT15}, is left open.} 




\sectS
\paragraph{Acknowledgements}
Richard Trefler was supported, in part, by an Individual Discovery Grant from the Natural Sciences and Engineering Research Council of Canada.
\sectS
\bibliographystyle{abbrv}
\bibliography{biblio}
\sectS
\end{document}